\documentclass[a4paper,11pt]{article}
\usepackage{fix-cm}
\usepackage{iftex}
\ifPDFTeX
  \pdfoutput=1
\fi
\makeatletter
\gdef\@fpheader{Prepared for submission to JHEP}
\gdef\@journal{jhep}

\newif\ifnotoc\notocfalse
\newif\ifemailadd\emailaddfalse
\newif\iftoccontinuous\toccontinuousfalse
\newif\ifnatbibsort\natbibsorttrue

\RequirePackage{amsmath}
\RequirePackage{amssymb}
\RequirePackage{epsfig}
\RequirePackage{graphicx}
\ifnatbibsort\RequirePackage[numbers,sort&compress]{natbib}\else\RequirePackage[numbers,compress]{natbib}\fi
\RequirePackage[colorlinks=true
  ,urlcolor=blue
  ,anchorcolor=blue
  ,citecolor=blue
  ,filecolor=blue
  ,linkcolor=blue
  ,menucolor=blue
  ,linktocpage=true
  ,pdfproducer=medialab
  ,pdfa=true
]{hyperref}

\def\@subheader{\@empty}
\def\@keywords{\@empty}
\def\@abstract{\@empty}
\def\@xtum{\@empty}
\def\@dedicated{\@empty}
\def\@arxivnumber{\@empty}
\def\@collaboration{\@empty}
\def\@collaborationImg{\@empty}
\def\@proceeding{\@empty}
\def\@preprint{\@empty}

\newcommand{\subheader}[1]{\gdef\@subheader{#1}}
\newcommand{\keywords}[1]{\if!\@keywords!\gdef\@keywords{#1}\else%
\PackageWarningNoLine{\jname}{Keywords already defined.\MessageBreak Ignoring last definition.}\fi}
\renewcommand{\abstract}[1]{\gdef\@abstract{#1}}
\newcommand{\dedicated}[1]{\gdef\@dedicated{#1}}
\newcommand{\arxivnumber}[1]{\gdef\@arxivnumber{#1}}
\newcommand{\proceeding}[1]{\gdef\@proceeding{#1}}
\newcommand{\xtumfont}[1]{\textsc{#1}}
\newcommand{\correctionref}[3]{\gdef\@xtum{\xtumfont{#1} \href{#2}{#3}}}
\newcommand\jname{JHEP}
\newcommand\acknowledgments{\section*{Acknowledgments}}
\newcommand\notoc{\notoctrue}
\newcommand\preprint[1]{\gdef\@preprint{\hfill #1}}

\newcommand\note[2][]{%
\if!#1!%
\stepcounter{footnote}\footnotetext{#2}%
\else%
{\renewcommand\thefootnote{#1}%
\footnotetext{#2}}%
\fi}

\newtoks\auth@toks
\renewcommand{\author}[2][]{%
  \if!#1!%
    \auth@toks=\expandafter{\the\auth@toks#2\ }%
  \else
    \auth@toks=\expandafter{\the\auth@toks#2$^{#1}$\ }%
  \fi
}

\newtoks\affil@toks\newif\ifaffil\affilfalse
\newcommand{\affiliation}[2][]{%
\affiltrue
  \if!#1!%
    \affil@toks=\expandafter{\the\affil@toks{\item[]#2}}%
  \else
    \affil@toks=\expandafter{\the\affil@toks{\item[$^{#1}$]#2}}%
  \fi
}

\newtoks\email@toks\newcounter{email@counter}%
\newcommand{\emailAdd}[1]{%
\emailaddtrue%
\ifnum\theemail@counter>0\email@toks=\expandafter{\the\email@toks, \@email{#1}}%
\else\email@toks=\expandafter{\the\email@toks\@email{#1}}%
\fi\stepcounter{email@counter}}
\newcommand{\@email}[1]{\href{mailto:#1}{\tt #1}}

\newcommand*\collaboration[1]{\gdef\@collaboration{#1}}
\newcommand*\collaborationImg[2][]{\gdef\@collaborationImg{#2}}

\newcommand\afterLogoSpace{\smallskip}
\newcommand\afterSubheaderSpace{\vskip3pt plus 2pt minus 1pt}
\newcommand\afterProceedingsSpace{\vskip21pt plus0.4fil minus15pt}
\newcommand\afterTitleSpace{\vskip23pt plus0.06fil minus13pt}
\newcommand\afterRuleSpace{\vskip23pt plus0.06fil minus13pt}
\newcommand\afterCollaborationSpace{\vskip3pt plus 2pt minus 1pt}
\newcommand\afterCollaborationImgSpace{\vskip3pt plus 2pt minus 1pt}
\newcommand\afterAuthorSpace{\vskip5pt plus4pt minus4pt}
\newcommand\afterAffiliationSpace{\vskip3pt plus3pt}
\newcommand\afterEmailSpace{\vskip16pt plus9pt minus10pt\filbreak}
\newcommand\afterXtumSpace{\par\bigskip}
\newcommand\afterAbstractSpace{\vskip16pt plus9pt minus13pt}
\newcommand\afterKeywordsSpace{\vskip16pt plus9pt minus13pt}
\newcommand\afterArxivSpace{\vskip3pt plus0.01fil minus10pt}
\newcommand\afterDedicatedSpace{\vskip0pt plus0.01fil}
\newcommand\afterTocSpace{\bigskip\medskip}
\newcommand\afterTocRuleSpace{\bigskip\bigskip}
\newlength{\affiliationsSep}
\newcommand\beforetochook{\pagestyle{myplain}\pagenumbering{roman}}

\DeclareFixedFont\trfont{OT1}{phv}{b}{sc}{11}

\renewcommand\maketitle{
\pagestyle{empty}
\thispagestyle{titlepage}
\setcounter{page}{0}
\noindent{\small\scshape\@fpheader}\@preprint\par
\afterLogoSpace
\if!\@subheader!\else\noindent{\trfont{\@subheader}}\fi
\afterSubheaderSpace
\if!\@proceeding!\else\noindent{\sc\@proceeding}\fi
\afterProceedingsSpace
{\LARGE\flushleft\sffamily\bfseries\@title\par}
\afterTitleSpace
\hrule height 1.5\p@%
\afterRuleSpace
\if!\@collaboration!\else
{\Large\bfseries\sffamily\raggedright\@collaboration}\par
\afterCollaborationSpace
\fi
\if!\@collaborationImg!\else
{\normalsize\bfseries\sffamily\raggedright\@collaborationImg}\par
\afterCollaborationImgSpace
\fi
{\bfseries\raggedright\sffamily\the\auth@toks\par}
\afterAuthorSpace
\ifaffil\begin{list}{}{%
\setlength{\leftmargin}{0.28cm}%
\setlength{\labelsep}{0pt}%
\setlength{\itemsep}{\affiliationsSep}%
\setlength{\topsep}{-\parskip}}
\itshape\small%
\the\affil@toks
\end{list}\fi
\afterAffiliationSpace
\ifemailadd 
\noindent\hspace{0.28cm}\begin{minipage}[l]{.9\textwidth}
\begin{flushleft}
\textit{E-mail:} \the\email@toks
\end{flushleft}
\end{minipage}
\else 
\PackageWarningNoLine{\jname}{E-mails are missing.\MessageBreak Plese use \protect\emailAdd\space macro to provide e-mails.}
\fi
\afterEmailSpace
\if!\@xtum!\else\noindent{\@xtum}\afterXtumSpace\fi
\if!\@abstract!\else\noindent{\renewcommand\baselinestretch{.9}\textsc{Abstract:}}\ \@abstract\afterAbstractSpace\fi
\if!\@keywords!\else\noindent{\textsc{Keywords:}} \@keywords\afterKeywordsSpace\fi
\if!\@arxivnumber!\else\noindent{\textsc{ArXiv ePrint:}} \href{https://arxiv.org/abs/\@arxivnumber}{\@arxivnumber}\afterArxivSpace\fi
\if!\@dedicated!\else\vbox{\small\it\raggedleft\@dedicated}\afterDedicatedSpace\fi
\ifnotoc\else
\iftoccontinuous\else\newpage\fi
\beforetochook\hrule
\tableofcontents
\afterTocSpace
\hrule
\afterTocRuleSpace
\fi
\setcounter{footnote}{0}
\pagestyle{myplain}\pagenumbering{arabic}
} 

\renewcommand{\baselinestretch}{1.1}\normalsize
\divide\@tempdima\baselineskip
\@tempcnta=\@tempdima
\skip\@mpfootins = \skip\footins
\renewcommand{\@dotsep}{10000}

\newcommand\ps@myplain{
\pagenumbering{arabic}
\renewcommand\@oddfoot{\hfill-- \thepage\ --\hfill}
\renewcommand\@oddhead{}}
\let\ps@plain=\ps@myplain

\newcommand\ps@titlepage{\renewcommand\@oddfoot{}\renewcommand\@oddhead{}}

\numberwithin{equation}{section}

\renewcommand\section{\@startsection{section}{1}{\z@}%
                                   {-3.5ex \@plus -1.3ex \@minus -.7ex}%
                                   {2.3ex \@plus.4ex \@minus .4ex}%
                                   {\normalfont\large\bfseries}}
\renewcommand\subsection{\@startsection{subsection}{2}{\z@}%
                                   {-2.3ex\@plus -1ex \@minus -.5ex}%
                                   {1.2ex \@plus .3ex \@minus .3ex}%
                                   {\normalfont\normalsize\bfseries}}
\renewcommand\subsubsection{\@startsection{subsubsection}{3}{\z@}%
                                   {-2.3ex\@plus -1ex \@minus -.5ex}%
                                   {1ex \@plus .2ex \@minus .2ex}%
                                   {\normalfont\normalsize\bfseries}}
\renewcommand\paragraph{\@startsection{paragraph}{4}{\z@}%
                                   {1.75ex \@plus1ex \@minus.2ex}%
                                   {-1em}%
                                   {\normalfont\normalsize\bfseries}}
\renewcommand\subparagraph{\@startsection{subparagraph}{5}{\z@}%
                                   {1.75ex \@plus1ex \@minus .2ex}%
                                   {-1em}%
                                   {\normalfont\normalsize\itshape}}

\def\fnum@figure{\textbf{\figurename\nobreakspace\thefigure}}
\def\fnum@table{\textbf{\tablename\nobreakspace\thetable}}

\long\def\@makecaption#1#2{%
  \vskip\abovecaptionskip
  \sbox\@tempboxa{\small #1. #2}%
  \ifdim \wd\@tempboxa >\hsize
    \small #1. #2\par
  \else
    \global \@minipagefalse
    \hb@xt@\hsize{\hfil\box\@tempboxa\hfil}%
  \fi
  \vskip\belowcaptionskip}

\renewenvironment{thebibliography}[1]{%
\begin{oldthebibliography}{#1}%
\small%
\raggedright%
\setlength{\itemsep}{5pt plus 0.2ex minus 0.05ex}%
}%
{%
\end{oldthebibliography}%
}
\makeatother

\newcommand{\ChineseAuthor}[1]{}
\ifPDFTeX
  \usepackage[T1]{fontenc}
  \usepackage{lmodern}
  \usepackage[utf8]{inputenc}
  \IfFileExists{CJKutf8.sty}{%
    \usepackage{CJKutf8}
    \renewcommand{\ChineseAuthor}[1]{ (##1)}
    \newenvironment{DocumentChineseNames}{\begin{CJK*}{UTF8}{gbsn}}{\end{CJK*}}
  }{\newenvironment{DocumentChineseNames}{}{}}
\else
  \usepackage[no-math]{fontspec}
  \IfFileExists{xeCJK.sty}{%
    \usepackage{xeCJK}
    \setCJKmainfont{FandolSong-Regular.otf}[BoldFont=FandolSong-Bold.otf]
    \setCJKsansfont{FandolHei-Regular.otf}[BoldFont=FandolHei-Bold.otf]
    \setCJKmonofont{FandolFang-Regular.otf}
    \renewcommand{\ChineseAuthor}[1]{ (##1)}
  }{}
  \newenvironment{DocumentChineseNames}{}{}
\fi
\usepackage{mathtools,bm,booktabs,tikz,flafter}
\makeatletter
\def\@fb@botlist{\@botlist}
\def\@fb@topbarrier{\suppressfloats[t]}
\def\FloatBarrier{\par\begingroup \let\@elt\relax

 \edef\@tempa{\@fb@botlist\@deferlist\@dbldeferlist}%
 \ifx\@tempa\@empty
 \else
    \ifx\@fltovf\relax 
       \if@firstcolumn
         \clearpage
       \else
         \null\newpage\FloatBarrier
       \fi
    \else
       \newpage \let\@fltovf\relax
       \FloatBarrier 
 \fi\fi \endgroup
 \@fb@topbarrier }

\makeatother

\usetikzlibrary{calc,positioning}
\allowdisplaybreaks
\notoc

\definecolor{allplusInk}{HTML}{26343D}
\definecolor{allplusBlue}{HTML}{276B8C}
\definecolor{allplusGrey}{HTML}{68767E}
\definecolor{allplusPale}{HTML}{F2F6F8}
\tikzset{
  allplus cut/.style={draw=allplusBlue,line width=1.0pt},
  allplus small/.style={font=\fontsize{9}{11}\selectfont,text=allplusInk},
  allplus heading/.style={font=\fontsize{10}{12}\selectfont\bfseries,
    text=allplusInk,anchor=west},
  allplus label/.style={font=\fontsize{9}{11}\selectfont,text=allplusGrey}
}
\newcommand{\allplusHexagon}[5][]{%
  \begin{scope}[shift={(#2,#3)}]
    \foreach \i in {1,...,6}{\coordinate (v\i) at ({180-60*\i}:.77);}
    \fill[allplusPale] (v1)--(v2)--(v3)--(v4)--(v5)--(v6)--cycle;
    #1%
    \draw[allplusInk,line width=.6pt] (v1)--(v2)--(v3)--(v4)--(v5)--(v6)--cycle;
    \foreach \a/\b in {#4}{\draw[allplus cut] (v\a)--(v\b);}
    \node[allplus small] at (0,-1.08) {$#5$};
  \end{scope}%
}

\title{All-plus Yang--Mills and gravity wavefunctions in de~Sitter space}
\author[1]{Song He\ChineseAuthor{何颂}}
\emailAdd{songhe@itp.ac.cn}
\author[2]{Jiajie Mei\ChineseAuthor{梅嘉杰}}
\emailAdd{j.mei@uva.nl}
\author[1]{Yuyu Mo\ChineseAuthor{莫裕宇}}
\emailAdd{moyuyu@itp.ac.cn}
\affiliation[1]{New Cornerstone Laboratory, Institute of Theoretical Physics,
Chinese Academy of Sciences, Beijing 100190, China}
\affiliation[2]{Institute of Physics, University of Amsterdam,
Amsterdam, 1098 XH, The Netherlands}

\abstract{
We present sewing representations for tree-level all-plus wavefunctions
in four-dimensional de Sitter space. For Yang--Mills theory, the
arbitrary-multiplicity formula is a sum over noncrossing polygon
dissections, with each cell carrying energy differences of a full
lower-point wavefunction. The absence of a generic total-energy pole
is manifest, while partial-energy residues are determined by the
lower-point factors. We derive this representation from self-dual Yang--Mills solutions.
For Einstein gravity, we propose an arbitrary-multiplicity formula
consisting of an alternating sum over labelled sewing trees plus a
local gravitational Chern--Simons term. This local term arises from the self-dual Einstein action and supplies
boundary data not fixed by internal-energy cuts. We derive a
generating formula for it from the boundary connection and give
explicit expressions through seven points.
}

\begin{document}
\begin{DocumentChineseNames}
\maketitle

\section{Introduction}
\label{sec:introduction}

The late-time wavefunction describes interacting fields in de Sitter
space and determines their boundary correlation
functions~\cite{MaldacenaPimentel2011,ArkaniHamedBootstrap2020}.
Its analytic structure connects these observables to scattering
amplitudes: flat-space amplitudes appear in total-energy
singularities~\cite{Raju2012}, while cosmological cutting rules relate
discontinuities to products of lower-point wavefunction
coefficients~\cite{GoodhewOptical2021,MelvillePajer2021,GoodhewCutting2021}.
For gauge fields and gravitons, extracting this information also
requires control over the gauge constraints and the boundary conditions
that define the wavefunction~\cite{BaumannSpinning2021}.

All-plus helicity configurations provide a particularly simple setting
for studying these structures. Their flat-space tree amplitudes vanish
for four or more external particles~\cite{Dixon2014}, whereas the
corresponding de Sitter wavefunction coefficients are generically
nonzero. Their partial-energy dependence and local boundary terms
therefore contain information beyond the flat-space limit. In our
recent work, we derived all-multiplicity scalar-QED wavefunctions from
Abelian Ward identities~\cite{HeQED2026} and reconstructed Yang--Mills
wavefunctions from cosmological cuts through six
points~\cite{HeCuts2026}. Here we study how the classical boundary-value
problem organizes the all-plus coefficients in Yang--Mills theory and
gravity.

Self-dual Yang--Mills solutions provide the classical starting
point~\cite{RoslySelivanov1997}. We compare the physical Dirichlet
problem with a formal self-dual reference problem having the same
linear, positive-helicity boundary source. The reference on-shell
action is at most cubic in this source. The difference between the self-dual and physical Green functions
factorizes, producing one energy-difference operation at each endpoint
of a sewn line. We derive this representation in Section~\ref{sec:constructive}. This Dirichlet
construction differs from the Neumann boundary problem studied
recently in Ref.~\cite{Heuveline2026}.

The Yang--Mills representation uses energy differences of full
lower-point coefficients.
Each compatible set of planar cuts contributes once, and there is no
additional uncut all-plus term. Noncrossing polygon dissections, or
equivalently proper faces of an associahedron, label the terms. We give
the explicit four-, five- and six-point formulas; at six points,
branching sewings first appear alongside chains. This organization
makes the simplification of the pole structure visible term by term.
Partial-energy poles and their successive residues are inherited from
the lower-point factors, while compatible cuts organize the simultaneous
singular limits. These features go beyond the vanishing flat-space
amplitude and connect to the general structure of cosmological
singularities~\cite{BaumannLinking2022}.

For Einstein gravity we propose an analogous sewing representation at
arbitrary multiplicity, built from full lower-point gravity coefficients
and a local gravitational Chern--Simons boundary term. The sum runs over
trees with distinctly labelled external legs, without a cyclic-order
restriction, and its signs alternate with the number of sewn edges.
This organization extends the simplification beyond conformally
invariant Yang--Mills theory. 
We give explicit four- and five-point relations using full lower-point
wavefunctions~\cite{BonifacioGraviton2023,MeiMo2026} and describe the
six-point sewing structure. Appendix~\ref{app:local-gravity} gives the
four- and five-point Chern--Simons kernels and a generating connection
expansion, including compact six- and seven-point tensors.

\paragraph{Note added.}
After our results had been obtained, Ref.~\cite{Nabavi2026} appeared,
presenting a diagrammatic construction of Yang--Mills wavefunctions
at arbitrary multiplicity. Its all-plus results overlap with those
presented here.

\section{All-plus Yang--Mills wavefunctions at all multiplicities}
\label{sec:ym-allplus}

We write $\Psi_n$ for the tree-level Yang--Mills wavefunction coefficients
of the physical Dirichlet problem, omitting the overall momentum
delta function.
Our coefficient convention is $\Psi_n=-S_{\mathrm{on\text{-}shell},n}$.
In this section the color order is $(1,\ldots,n)$, and all unmarked
external helicities are positive. For a consecutive set of labels $I$,
write
\begin{equation}
 \bm P_I=\sum_{i\in I}\bm k_i,\qquad
 p_I=|\bm P_I|,\qquad k_i=|\bm k_i|,\qquad
 D_p f=f(p)-f(-p).
 \label{eq:ym-letter-disc}
\end{equation}
Here $p$ is the radial-energy argument of a specified leg. The spatial
momentum, polarization and spatial contractions of that leg are held
fixed. 

Both ends of a sewn line have negative helicity relative to their own
outgoing momenta. We use $\epsilon_+(\bm P)$ for $(-P)^-$ and
$\epsilon_-(\bm P)$ for $P^-$. The sewing factor is $1/(2p)$ and
there is no helicity sum. Thus each line applies one $D_p$ to each of
the two wavefunctions it joins.

To specify the sum, place the external legs on the edges of an $n$-gon.
A chord $(a,b)$ denotes the channel $I|I^c$, with
$I=\{a,a+1,\ldots,b-1\}$. A set $C$ of noncrossing chords divides the
polygon into cells $f$. Associate a full wavefunction coefficient to
each cell and define
\begin{equation}
 B_C=\left(\prod_{e\in C}\frac{1}{2p_e}\right)
 \prod_{f\in\mathrm{cells}(C)}
 \left[\left(\prod_{e\in C\cap\partial f}D_{e,f}\right)
       \Psi_{|f|}\right].
 \label{eq:ym-letter-weight}
\end{equation}
The number $|f|$ includes the sewn legs, and their order is inherited
from the polygon. The subscript $f$ on $D_{e,f}$ specifies which
wavefunction it acts on.

The sewing diagrams are the planar duals of these polygon dissections.
Place a vertex in each cell, join neighbouring vertices across their
common chord, and attach an external leg across each boundary edge.
A cell with $m$ sides gives a node carrying the full $\Psi_m$; each chord
gives a sewn edge with weight $1/(2p)$ and one $D_p$ at each endpoint.
Thus $r$ chords give a tree with $r+1$ wavefunction nodes.
Figures~\ref{allplus:fig:n} and~\ref{allplus:fig:ym6} use these two
equivalent representations of the same terms.

Let $\mathcal N_{n,r}$ denote the sets of $r$ noncrossing chords.
The all-multiplicity sewing representation is
\begin{equation}
 \boxed{\displaystyle
 \Psi_n^{+\cdots+}
 =-\sum_{r=1}^{n-3}\ \sum_{C\in\mathcal N_{n,r}}B_C,
 \qquad n\ge4.}
 \label{eq:ym-letter-alln}
\end{equation}
Each nonempty dissection occurs once, with the same overall minus sign.
The coefficients associated with its cells have at most $n-1$ legs
and generally mixed helicities; they include all exchange and contact
contributions, while no independent uncut $n$-point term is required.

Equation~\eqref{eq:ym-letter-alln} is derived in
Section~\ref{sec:constructive}. With a linear positive-helicity
boundary source, the self-dual reference problem has no all-plus
coefficient beyond three points. Its Green function differs from the
physical one by a factorized term. Expanding in this difference
produces the sewn products, while the vanishing reference coefficient
fixes their common overall minus sign.

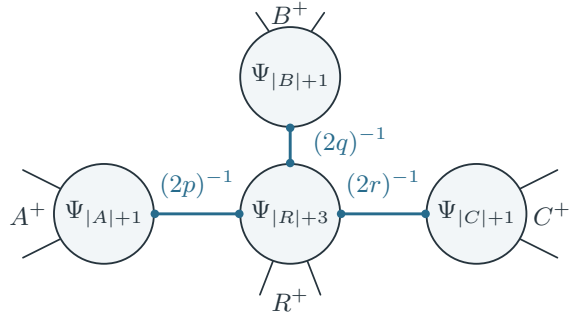
\begin{figure}[!htbp]
\centering
\begin{tikzpicture}[x=1cm,y=1cm,line cap=round,line join=round,
  every node/.style={inner sep=1.5pt},
  ym kernel/.style={circle,draw=allplusInk,fill=allplusPale,
    line width=.7pt,minimum size=1.32cm,inner sep=1.5pt,
    font=\fontsize{9.5}{11}\selectfont,text=allplusInk},
  ym small node/.style={circle,draw=allplusInk,fill=allplusPale,
    line width=.7pt,minimum size=.87cm,inner sep=2pt,
    font=\fontsize{10}{12}\selectfont,text=allplusInk},
  ym leg/.style={draw=allplusInk,line width=.65pt},
  ym edge/.style={draw=allplusBlue,line width=1.1pt},
  ym label/.style={font=\fontsize{9.5}{11}\selectfont,text=allplusInk},
  ym weight/.style={font=\fontsize{9.5}{11}\selectfont,text=allplusBlue}]
  \path[use as bounding box] (0,-.25) rectangle (14.8,4.85);
  \node[allplus heading] at (.05,4.63) {(a) A single sewn line};
  \node[allplus heading] at (7.65,4.63) {(b) A three-cut contribution};
  \draw[allplusGrey!25,line width=.4pt] (7.20,.03)--(7.20,4.35);

  \node[ym label] at (3.50,3.91)
    {$\displaystyle\Delta G^{ij}
      =\frac{\epsilon_P^{-i}\epsilon_P^{+j}}{2p}\,d_p(z)d_p(t)$};
  \node[ym label] at (3.50,3.25) {$d_p(z)=e^{-pz}-e^{pz}$};
  \draw[->,allplusGrey,line width=.65pt] (3.50,2.96)--(3.50,2.40);
  \node[ym small node] (L) at (1.78,1.52) {$\Psi_L$};
  \node[ym small node] (R) at (5.22,1.52) {$\Psi_R$};
  \draw[ym leg] (L.145)--(.57,2.22) (L.215)--(.57,.82);
  \draw[ym leg] (R.35)--(6.43,2.22) (R.325)--(6.43,.82);
  \node[ym label] at (.61,1.52) {$\vdots$};
  \node[ym label] at (6.39,1.52) {$\vdots$};
  \draw[ym edge] (L.east)--(R.west);
  \foreach \p in {L.east,R.west}
    {\fill[allplusBlue] (\p) circle[radius=1.6pt];}
  \node[ym weight] at (3.50,1.95) {$(2p)^{-1}$};
  \node[ym label] at (2.38,2.17) {$P^-$};
  \node[ym label] at (4.62,2.17) {$(-P)^-$};
  \node[ym weight] at (2.38,1.08) {$D_p$};
  \node[ym weight] at (4.62,1.08) {$D_p$};
  \node[ym label] at (3.50,.20)
    {$\displaystyle\frac{(D_p\Psi_L)(D_p\Psi_R)}{2p}$};

  \node[ym kernel] (A) at (8.72,1.47) {$\Psi_{|A|+1}$};
  \node[ym kernel] (B) at (11.18,3.28) {$\Psi_{|B|+1}$};
  \node[ym kernel] (C) at (13.64,1.47) {$\Psi_{|C|+1}$};
  \node[ym kernel] (V) at (11.18,1.47) {$\Psi_{|R|+3}$};
  \draw[ym edge] (A.east)--(V.west) (B.south)--(V.north)
    (V.east)--(C.west);
  \foreach \p in {A.east,V.west,B.south,V.north,V.east,C.west}
    {\fill[allplusBlue] (\p) circle[radius=1.6pt];}
  \node[ym weight] at (9.95,1.86) {$(2p)^{-1}$};
  \node[ym weight,anchor=west] at (11.42,2.42) {$(2q)^{-1}$};
  \node[ym weight] at (12.41,1.86) {$(2r)^{-1}$};
  \draw[ym leg] (A.150)--(7.65,2.05) (A.210)--(7.65,.89);
  \node[ym label] at (7.72,1.47) {$A^+$};
  \draw[ym leg] (B.115)--(10.73,4.13) (B.65)--(11.63,4.13);
  \node[ym label] at (11.18,4.13) {$B^+$};
  \draw[ym leg] (C.30)--(14.71,2.05) (C.330)--(14.71,.89);
  \node[ym label] at (14.64,1.47) {$C^+$};
  \draw[ym leg] (V.250)--(10.78,.40) (V.290)--(11.58,.40);
  \node[ym label] at (11.18,.33) {$R^+$};
\end{tikzpicture}
\caption{Yang--Mills sewing rules. (a) The factorized difference
$\Delta G=G_{\mathrm{SD}}-G_{\mathrm{phys}}$ gives one $D_p$ at each
endpoint and weight $1/(2p)$; both endpoint helicities refer to outgoing
momenta. (b) A three-cut term with external sets in cyclic order
$(A,B,C,R)$, $|A|,|B|,|C|\ge2$, and $R$ possibly empty.
Here $p=|\bm P_A|$, $q=|\bm P_B|$, $r=|\bm P_C|$, and
$n=|A|+|B|+|C|+|R|$. Blue dots apply energy differences to full
wavefunctions; all sewn legs have negative helicity.
For $A=\{1,2\}$, $B=\{3,4\}$, $C=\{5,6\}$ and $R=\varnothing$,
the central node carries $D_pD_qD_r\Psi_3^{---}$ and is dual to the
shaded triangle in Fig.~\ref{allplus:fig:ym6}.}
\label{allplus:fig:n}
\end{figure}

At four points the two allowed cuts give
\begin{equation}
 \Psi_4^{++++}
 =-\sum_{I=12,23}
 \frac{D_{p_I}\Psi_3(I,(-P_I)^-)\,
       D_{p_I}\Psi_3(P_I^-,I^c)}{2p_I}.
 \label{eq:ym-letter-four}
\end{equation}
For $I=23$ the complement is ordered as $(4,1)$.
At five points there are two cyclic representatives: a single cut
joining three- and four-point wavefunctions, and a two-cut chain of
three-point wavefunctions. Writing $P=P_{12}$, $Q=P_{45}$,
$p=p_{12}$ and $q=p_{45}$, the result is
\begin{equation}
 \begin{split}
 \Psi_5^{+++++}=-\operatorname{Cyc}_5\Bigg[&
 \frac{D_p\Psi_3(1,2,(-P)^-)\,
       D_p\Psi_4(P^-,3,4,5)}{2p}\\
 &+\frac{D_p\Psi_3(1,2,(-P)^-)\,
       D_pD_q\Psi_3(P^-,3,Q^-)\,
       D_q\Psi_3((-Q)^-,4,5)}{4pq}\Bigg].
 \end{split}
 \label{eq:ym-letter-five}
\end{equation}
Here $\operatorname{Cyc}_n F=\sum_{j=0}^{n-1}\rho^j F$,
where $\rho:i\mapsto i+1$ rotates every label, including those in the
channel momenta. The four-point input is the full $-+++$ coefficient.

The six-point result has nine one-cut, twenty-one two-cut and fourteen
three-cut terms. It is convenient to write $B_{I_1,\ldots,I_r}$ for
the cut set specified by the particle channels $I_1,\ldots,I_r$;
$I$ and $I^c$ specify the same cut. The complete expression is
\begin{equation}
 \begin{split}
 \Psi_6^{+\cdots+}=-\operatorname{Cyc}_6\Big[&
 B_{12}+\tfrac12 B_{123}\\
 &+B_{12,123}+B_{12,56}+B_{12,345}+\tfrac12 B_{12,45}\\
 &+B_{12,123,56}+\tfrac12 B_{12,123,45}
   +\tfrac13 B_{12,56,34}+\tfrac12 B_{12,345,45}\Big].
 \end{split}
 \label{eq:ym-letter-six}
\end{equation}
The three lines group terms by cut number. Factors of $1/2$ and $1/3$
remove repetitions of the shorter cyclic orbits. All ten
representatives are drawn in Fig.~\ref{allplus:fig:ym6}. The one-cut
terms include both $3\times5$ and $4\times4$ products. Among the
fully cut terms are two stars, whose central coefficient is
$D_pD_qD_r\Psi_3^{---}$. This coefficient is evaluated with
independent energy arguments before taking the three differences.

\begin{figure}[!htbp]
\centering
\begin{tikzpicture}[x=1cm,y=1cm,line cap=round,line join=round]
  \path[use as bounding box] (-.2,-6.95) rectangle (14.6,1.3);
  \node[allplus heading] at (0,1.17) {One cut};
  \node[allplus label,anchor=east] at (14.5,1.17) {$6+3=9$};
  \draw[allplusGrey!25,line width=.4pt] (0,.97)--(14.5,.97);
  \allplusHexagon{1.5}{0}{1/3}{\{12\}\quad[6]}
  \allplusHexagon{5.1}{0}{1/4}{\{123\}\quad[3]}
  \begin{scope}[shift={(8.7,0)}]
    \foreach \i in {1,...,6}{\coordinate (w\i) at ({180-60*\i}:.59);}
    \draw[allplusGrey,line width=.5pt] (w1)--(w2)--(w3)--(w4)--(w5)--(w6)--cycle;
    \foreach \i in {1,...,6}{
      \node[allplus label] at ({150-60*\i}:.82) {$\i$};
    }
    \node[allplus label] at (0,-1.08) {edge labels};
  \end{scope}
  \node[allplus small,align=center] at (12.3,.15)
    {$9+21+14=44$\\[5pt]$[m]$: $m$ distinct rotations};
  \node[allplus heading] at (0,-1.63) {Two cuts};
  \node[allplus label,anchor=east] at (14.5,-1.63) {$6+6+6+3=21$};
  \draw[allplusGrey!25,line width=.4pt] (0,-1.83)--(14.5,-1.83);
  \allplusHexagon{1.5}{-2.8}{1/3,1/4}{\{12,123\}\quad[6]}
  \allplusHexagon{5.1}{-2.8}{1/3,1/5}{\{12,56\}\quad[6]}
  \allplusHexagon{8.7}{-2.8}{1/3,3/6}{\{12,345\}\quad[6]}
  \allplusHexagon{12.3}{-2.8}{1/3,4/6}{\{12,45\}\quad[3]}
  \node[allplus heading] at (0,-4.43) {Three cuts};
  \node[allplus label,anchor=east] at (14.5,-4.43) {$6+3+2+3=14$};
  \draw[allplusGrey!25,line width=.4pt] (0,-4.63)--(14.5,-4.63);
  \allplusHexagon{1.5}{-5.6}{1/3,1/4,1/5}{\{12,123,56\}\quad[6]}
  \allplusHexagon{5.1}{-5.6}{1/3,1/4,4/6}{\{12,123,45\}\quad[3]}
  \allplusHexagon[{\fill[allplusBlue!20] (v1)--(v3)--(v5)--cycle;
    \node[font=\fontsize{8}{9}\selectfont,text=allplusBlue] at (.06,0)
      {$\Psi_3^{---}$};}]{8.7}{-5.6}{1/3,1/5,3/5}{\{12,56,34\}\quad[2]}
  \allplusHexagon{12.3}{-5.6}{1/3,3/6,4/6}{\{12,345,45\}\quad[3]}
\end{tikzpicture}
\caption{The ten six-point cyclic representatives. Each blue chord is dual
to a sewn edge and each polygonal cell denotes a full wavefunction coefficient.
All polygons share the edge labelling shown at the top; bracketed
numbers count distinct cyclic rotations. The sets below the diagrams
label particle channels.
The shaded central triangle of $\{12,56,34\}$ is dual to the central
node in Fig.~\ref{allplus:fig:n}(b) and carries
$D_pD_qD_r\Psi_3^{---}$, with three independent energy arguments.}
\label{allplus:fig:ym6}
\end{figure}
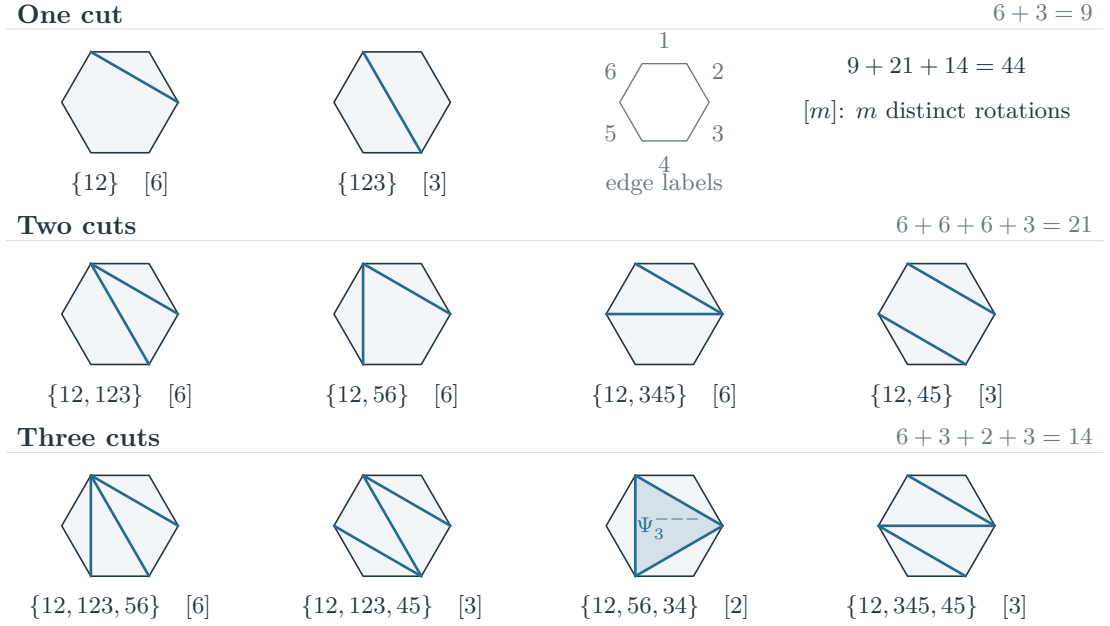

The energy-pole structure of Eq.~\eqref{eq:ym-letter-alln} is simpler
than that of individual bulk diagrams. Every sewn factor depends on a
proper subset of external energies together with its channel energies,
so no term has a pole at generic $E_T=\sum_{i=1}^n k_i=0$.
Partial-energy singularities remain: their poles and residues are
inherited from the lower-point coefficients. Successive residues can
therefore be studied by taking the corresponding limits of these
factors. Compatible dissections specify which factors occur together,
organizing the possible intersections of singularities; numerator
cancellations can further remove a residue at such an
intersection~\cite{BaumannLinking2022}.

There is a simple combinatorial description of the same sum. A set of
$r$ noncrossing chords labels a codimension-$r$ face of the
$(n-3)$-dimensional associahedron~\cite{ABHY2018}. Thus
Eq.~\eqref{eq:ym-letter-alln} is indexed by all its nonempty proper
faces. Writing $\mathcal A_d$ for the $d$-dimensional associahedron,
the face associated with $C$ factorizes as
$F_C\simeq\prod_{f\in\mathrm{cells}(C)}\mathcal A_{|f|-3}$,
with one factor for each cell, or equivalently each wavefunction node
of the dual tree. The number of terms at each cut order is the dissection
number~\cite{Gaiffi2014}
\begin{equation}
 N_{n,r}^{\mathrm{YM}}
 =\frac{1}{r+1}\binom{n-3}{r}\binom{n+r-1}{r},\qquad
 N_n^{\mathrm{YM}}=s_{n-2}-1,
 \label{eq:ym-letter-faces}
\end{equation}
where $s_0,s_1,\ldots=1,1,3,11,45,\ldots$ are the little
Schr\"oder, or super-Catalan, numbers~\cite{vonBellYip2020}.
The subtraction removes the empty cut set, leaving one term for each
proper face in the sewing representation.

\FloatBarrier
\section{All-plus gravity wavefunctions}
\label{sec:gr-allplus-short}

We propose an arbitrary-multiplicity sewing representation for all-plus
wavefunctions in Einstein gravity, consisting of an alternating sum over
labelled trees and a local gravitational Chern--Simons term. All distinct
external-label assignments contribute, without the cyclic-order
restriction of Yang--Mills theory.

We use $\Psi_n$ for the gravity coefficients in this section and the
same energy difference $D_p$ as in Eq.~\eqref{eq:ym-letter-disc}.
Both sewn legs again have negative helicity relative to their
outgoing momenta. We use $\xi^\pm_{ij}=\epsilon^\pm_i\epsilon^\pm_j$,
with $\epsilon_+\cdot\epsilon_-=1$ and
$\xi^-_{ij}(-\bm P)\xi^{-\,ij}(\bm P)=1$.
All lower-point coefficients use this normalization; the sewn-edge
weight is $2/p^3$.
We use the exponential boundary metric $g^{(0)}=e^\gamma$, with the
same source variable and boundary prescription in every lower-point
coefficient.

\subsection{A general sewing organization}
\label{sec:gr-pattern}
Let $\mathcal T_{n,r}$ denote the connected trees with $n$ distinctly
labelled external leaves, $r$ internal edges, and internal vertices of
valence at least three. No planarity or cyclic-order restriction is
imposed. Each distinct tree, up to graph isomorphisms preserving the
external labels, occurs once. Cutting an internal edge $e$ determines
a bipartition $I_e|I_e^c$, with
$p_e=|\sum_{i\in I_e}\bm k_i|$. Associate a full gravity coefficient
to each internal vertex and define
\begin{equation}
 B_T^{\mathrm{GR}}=
 \left(\prod_{e\in E(T)}\frac{2}{p_e^3}\right)
 \prod_{v\in V(T)}
 \left[\left(\prod_{\substack{e\in E(T)\\ e\ni v}}D_{e,v}\right)
       \Psi_{m_v}\right].
 \label{eq:gr-letter-weight}
\end{equation}
Here $V(T)$ is the set of internal vertices, $m_v$ counts all legs
incident on $v$, including sewn legs, and $D_{e,v}$ acts on the energy
argument of the corresponding endpoint. Every external leg has
positive helicity and every sewn endpoint has negative helicity.
There is no helicity sum on these sewn edges. Energy differences act
on the complete radial-mode dependence, including energy-dependent
numerators, with spatial momenta and polarization tensors held fixed.
The proposed general form is
\begin{equation}
 \boxed{\displaystyle
 \Psi_n^{+\cdots+}
 =\frac14\operatorname{CS}_n^{+\cdots+}
  +\sum_{r=1}^{n-3}(-1)^{r+1}
   \sum_{T\in\mathcal T_{n,r}}B_T^{\mathrm{GR}},\qquad n\ge4.}
 \label{eq:gr-letter-alln}
\end{equation}
The signs alternate with the number of sewn edges. Since
$\sum_v(m_v-2)=n-2$, each vertex in a nontrivial tree has
$m_v\le n-1$. The inputs are therefore full lower-point coefficients,
generally with mixed helicities, including all their exchange and
contact contributions. Vertices with no external leaves are allowed. We work at generic
nonzero channel momenta and use the same radial continuation and
boundary prescription in every factor. The term
$\tfrac14\operatorname{CS}_n$ is a local boundary contribution,
defined at every multiplicity in Appendix~\ref{app:local-gravity}.

\subsection{Four- and five-point examples}

We illustrate the general representation at four and five points.
At four points, the three bipartitions each contribute a product of
three-point coefficients, giving
\begin{equation}
 \Psi_4^{++++}
 =2\sum_{ij|lm}
 \frac{D_p\Psi_3(i,j,(-P)^-)\,
       D_p\Psi_3(P^-,l,m)}{p^3}
 +\frac14\operatorname{CS}_4^{++++},
 \label{eq:gr-short-four}
\end{equation}
where $ij|lm=12|34,13|24,14|23$, $P=P_{ij}$ and $p=p_{ij}$.
The three sewn terms and the local contribution are shown explicitly
in Fig.~\ref{allplus:fig:gravity}(a).

\begin{samepage}
The five-point formula involves the same two types of sewing as
Eq.~\eqref{eq:ym-letter-five}, now summed over all distinct external
labels:\par
\begin{equation}
 \begin{split}
 \Psi_5^{+++++}
 ={}&2\sum_{|I|=2}
 \frac{D_{p_I}\Psi_3(I,(-P_I)^-)\,
       D_{p_I}\Psi_4(P_I^-,I^c)}{p_I^3}\\
 &-4\sum_{\{I,J\}}
 \frac{D_{p_I}\Psi_3(I,(-P_I)^-)\,
       D_{p_I}D_{p_J}\Psi_3(P_I^-,m,P_J^-)\,
       D_{p_J}\Psi_3((-P_J)^-,J)}{p_I^3p_J^3}\\
 &+\frac14\operatorname{CS}_5^{+++++}.
 \end{split}
 \label{eq:gr-short-five}
\end{equation}
\end{samepage}
The first sum contains the ten two-element subsets of $[5]$.
The second contains the fifteen unordered pairs $\{I,J\}$ of
disjoint two-element subsets; $m$ is the remaining external leg.
In particular, exchanging $I$ and $J$ does not give an additional
term. The four-point input is the complete $-+++$ coefficient,
including exchanges and contacts. Figure~\ref{allplus:fig:gravity}(b)
shows one representative of each sum. The relative minus sign of
the two-cut term combines with the exchange contributions already
present in the full four-point input. The two sums thus reorganize
the bulk diagrams into products of full wavefunction coefficients.

The local term is essential to these identities. Internal-energy cuts
determine the sewn contributions but cannot detect a functional built
only from the boundary metric and spatial derivatives. In the fixed
source $g^{(0)}=e^\gamma$, the missing four- and five-point terms are
$\tfrac14\operatorname{CS}_4$ and $\tfrac14\operatorname{CS}_5$.
Changing the nonlinear metric variable would change their individual
kernels, so the same source convention must be used in every factor.
The metric variation of the gravitational Chern--Simons functional is
proportional to the Cotton tensor, which also governs the boundary
response of self-dual gravity~\cite{MansiPetkouTagliabue2009,deHaro2009}.

Equations~\eqref{eq:gr-short-four} and \eqref{eq:gr-short-five}
make the absence of a generic total-energy pole explicit. Their sewn
factors have only lower-point energy denominators, while the local
terms have no energy poles. The surviving partial-energy singularities
and their residues are consequently determined by the lower-point
factors. The local completion carries additional information fixed by
the gravitational boundary problem. Its explicit kernels are given in
Appendix~\ref{app:local-gravity}.

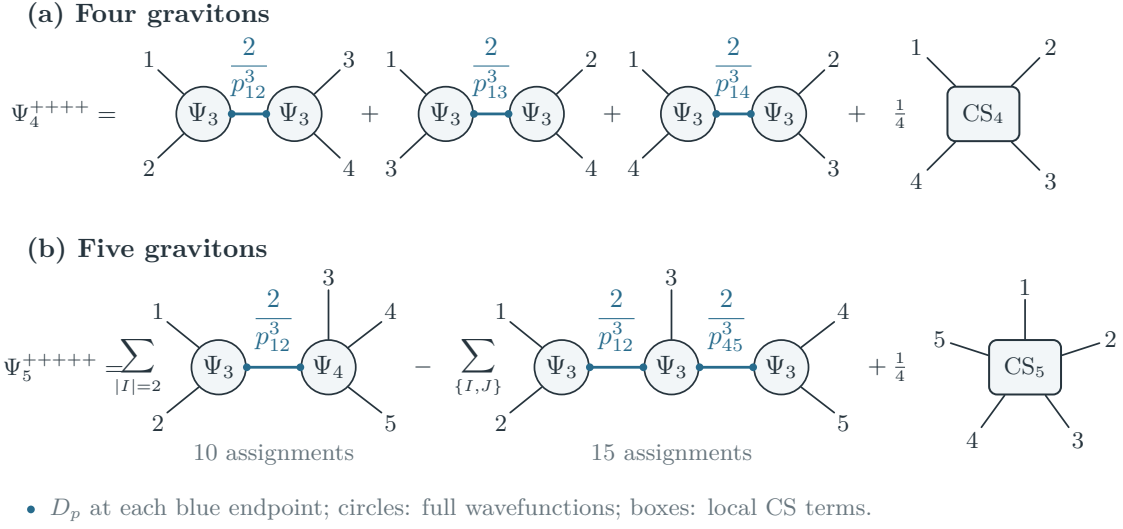
\begin{figure}[!htbp]
\centering
\begin{tikzpicture}[x=1cm,y=1cm,line cap=round,line join=round,
  every node/.style={inner sep=1.5pt},
  gr kernel/.style={circle,draw=allplusInk,fill=allplusPale,
    line width=.65pt,minimum size=.72cm,inner sep=0pt,
    font=\fontsize{10}{12}\selectfont,text=allplusInk},
  gr leg/.style={draw=allplusInk,line width=.65pt},
  gr edge/.style={draw=allplusBlue,line width=1.05pt},
  gr label/.style={font=\fontsize{9}{11}\selectfont,text=allplusInk},
  gr weight/.style={font=\fontsize{10}{12}\selectfont,text=allplusBlue},
  gr heading/.style={font=\fontsize{10}{12}\selectfont\bfseries,
    text=allplusInk,anchor=west},
  gr local/.style={rectangle,rounded corners=3pt,
    draw=allplusInk,fill=allplusPale,line width=.75pt,
    minimum width=.95cm,minimum height=.72cm,inner sep=2pt,
    font=\fontsize{9}{11}\selectfont,text=allplusInk}]
  \path[use as bounding box] (-.15,-.12) rectangle (14.65,6.95);
  \node[gr heading] at (.05,6.75) {(a) Four gravitons};
  \node[gr label] at (.60,5.45) {$\Psi_4^{++++}=$};

  \foreach \x/\a/\b/\c/\d/\p in
      {2.45/1/2/3/4/12,5.65/1/3/2/4/13,8.85/1/4/2/3/14}{
    \begin{scope}[shift={(\x,5.45)}]
      \node[gr kernel] (L) at (0,0) {$\Psi_3$};
      \node[gr kernel] (R) at (1.20,0) {$\Psi_3$};
      \draw[gr leg] (L.135)--(-.60,.58) node[gr label,above left=-1pt] {$\a$};
      \draw[gr leg] (L.225)--(-.60,-.58) node[gr label,below left=-1pt] {$\b$};
      \draw[gr leg] (R.45)--(1.80,.58) node[gr label,above right=-1pt] {$\c$};
      \draw[gr leg] (R.315)--(1.80,-.58) node[gr label,below right=-1pt] {$\d$};
      \draw[gr edge] (L.east)--(R.west);
      \fill[allplusBlue] (L.east) circle[radius=1.45pt];
      \fill[allplusBlue] (R.west) circle[radius=1.45pt];
      \node[gr weight] at (.60,.61) {$\dfrac{2}{p_{\p}^{3}}$};
    \end{scope}
  }
  \node[gr label] at (4.65,5.45) {$+$};
  \node[gr label] at (7.85,5.45) {$+$};
  \node[gr label] at (11.08,5.45) {$+$};
  \node[gr label] at (11.65,5.45) {$\frac14$};
  \node[gr local] (CS4) at (12.75,5.45) {$\mathrm{CS}_4$};
  \foreach \ang/\lab in {135/1,45/2,315/3,225/4}{
    \draw[gr leg] (CS4.\ang)--++(\ang:.52)
      node[gr label,anchor={\ang+180}] {$\lab$};
  }

  \node[gr heading] at (.05,3.65) {(b) Five gravitons};
  \node[gr label] at (.60,2.10) {$\Psi_5^{+++++}=$};
  \node[gr label] at (1.58,2.10) {$\displaystyle\sum_{|I|=2}$};
  \node[gr kernel] (S3) at (2.65,2.10) {$\Psi_3$};
  \node[gr kernel] (S4) at (4.10,2.10) {$\Psi_4$};
  \draw[gr leg] (S3.135)--(1.97,2.70) node[gr label,above left=-1pt] {$1$};
  \draw[gr leg] (S3.225)--(1.97,1.50) node[gr label,below left=-1pt] {$2$};
  \draw[gr leg] (S4.90)--(4.10,3.12) node[gr label,above] {$3$};
  \draw[gr leg] (S4.45)--(4.80,2.70) node[gr label,above right=-1pt] {$4$};
  \draw[gr leg] (S4.315)--(4.80,1.50) node[gr label,below right=-1pt] {$5$};
  \draw[gr edge] (S3.east)--(S4.west);
  \fill[allplusBlue] (S3.east) circle[radius=1.45pt];
  \fill[allplusBlue] (S4.west) circle[radius=1.45pt];
  \node[gr weight] at (3.375,2.72) {$\dfrac{2}{p_{12}^{3}}$};
  \node[gr label,text=allplusGrey] at (3.375,.94) {10 assignments};

  \node[gr label] at (5.35,2.10) {$-$};
  \node[gr label] at (6.08,2.10) {$\displaystyle\sum_{\{I,J\}}$};
  \node[gr kernel] (T1) at (7.18,2.10) {$\Psi_3$};
  \node[gr kernel] (T2) at (8.63,2.10) {$\Psi_3$};
  \node[gr kernel] (T3) at (10.08,2.10) {$\Psi_3$};
  \draw[gr leg] (T1.135)--(6.50,2.70) node[gr label,above left=-1pt] {$1$};
  \draw[gr leg] (T1.225)--(6.50,1.50) node[gr label,below left=-1pt] {$2$};
  \draw[gr leg] (T2.90)--(8.63,3.12) node[gr label,above] {$3$};
  \draw[gr leg] (T3.45)--(10.78,2.70) node[gr label,above right=-1pt] {$4$};
  \draw[gr leg] (T3.315)--(10.78,1.50) node[gr label,below right=-1pt] {$5$};
  \draw[gr edge] (T1.east)--(T2.west) (T2.east)--(T3.west);
  \foreach \v in {T1.east,T2.west,T2.east,T3.west}
    {\fill[allplusBlue] (\v) circle[radius=1.45pt];}
  \node[gr weight] at (7.905,2.72) {$\dfrac{2}{p_{12}^{3}}$};
  \node[gr weight] at (9.355,2.72) {$\dfrac{2}{p_{45}^{3}}$};
  \node[gr label,text=allplusGrey] at (8.63,.94) {15 assignments};

  \node[gr label] at (11.35,2.10) {$+$};
  \node[gr label] at (11.65,2.10) {$\frac14$};
  \node[gr local] (CS5) at (13.30,2.10) {$\mathrm{CS}_5$};
  \foreach \ang/\lab in {90/1,18/2,306/3,234/4,162/5}{
    \draw[gr leg] (CS5.\ang)--++(\ang:.52)
      node[gr label,anchor={\ang+180}] {$\lab$};
  }
  \fill[allplusBlue] (.17,.22) circle[radius=1.45pt];
  \node[gr label,anchor=west,text=allplusGrey] at (.36,.22)
    {$D_p$ at each blue endpoint; circles: full wavefunctions; boxes: local CS terms.};
\end{tikzpicture}
\caption{Four- and five-point gravity wavefunctions. A blue endpoint
applies $D_p$ to the incident full wavefunction; rounded boxes denote
local Chern--Simons terms. External legs have positive helicity and
sewn legs have negative helicity. At five points, $I$ has two labels
and $\{I,J\}$ is an unordered pair of disjoint two-label sets.}
\label{allplus:fig:gravity}
\end{figure}

\subsection{Higher multiplicity and counting}
At six points, the proposed representation \eqref{eq:gr-letter-alln} reads
\begin{equation}
 \begin{split}
 \Psi_6^{+\cdots+}
 ={}&\frac14\operatorname{CS}_6^{+\cdots+}
 +\sum_{T\in\mathcal T_{6,1}}B_T^{\mathrm{GR}}
 -\sum_{T\in\mathcal T_{6,2}}B_T^{\mathrm{GR}}
 +\sum_{T\in\mathcal T_{6,3}}B_T^{\mathrm{GR}}.
 \end{split}
 \label{eq:gr-letter-six}
\end{equation}
There are $25$ one-edge, $105$ two-edge and $105$ three-edge trees.
The one-edge terms comprise $15$ sewings of three- and five-point
coefficients and $10$ sewings of two four-point coefficients. The
two-edge terms contain one four-point and two three-point coefficients.
The three-edge terms contain four three-point coefficients and
comprise $90$ chains and $15$ stars. Each star has a central
$D_pD_qD_r\Psi_3^{---}$ coefficient with no external
legs; its three energy arguments are independent until the
differences have been taken.

\begin{samepage}
These trees have labelled external legs and internal vertices of
valence at least three, with no cyclic ordering. Their counts through
eight points are listed in Table~\ref{tab:allplus-counts}~\cite{Kampf2020},
together with the single local boundary contribution in
Eq.~\eqref{eq:gr-letter-alln}.\par
\end{samepage}

\begin{table}[!htbp]
\centering
\begin{tabular}{lrrrrr}
\toprule
$n$ & $4$ & $5$ & $6$ & $7$ & $8$\\
\midrule
YM sewn terms & $2$ & $10$ & $44$ & $196$ & $902$\\
GR terms, including local & $4$ & $26$ & $236$ & $2752$ & $39208$\\
\bottomrule
\end{tabular}
\caption{Counts with labelled external legs. The YM row fixes their cyclic
order and excludes the empty cut set. The GR row allows all internal
valences and counts the sewn labelled trees together with the single
local Chern--Simons boundary contribution.}
\label{tab:allplus-counts}
\end{table}


\FloatBarrier
\section{Derivation from self-dual solutions}
\label{sec:constructive}

After Euclidean continuation, conformal invariance maps the
Yang--Mills problem to the flat half-space $z\geq0$.
With $F_{\mu\nu}[A]=\partial_\mu A_\nu-\partial_\nu A_\mu-ig[A_\mu,A_\nu]$,
the action in axial gauge $A_z=0$ can be written as
\begin{equation}
 S[A]=\frac14\int d^4x\,\operatorname{tr}\!\left(F_{\mu\nu}F^{\mu\nu}\right)
 =\frac12\int dz\,d^3x\,\operatorname{tr}
 \left(F_{zi}+\frac12\epsilon_i{}^{jk}F_{jk}\right)^2
 +C_\partial[A(0)],
 \label{eq:construct-square}
\end{equation}
where $C_\partial$ is the boundary Chern--Simons functional,
\begin{equation}
 C_\partial[A]=\int d^3x\,\epsilon^{ijk}\operatorname{tr}
 \left(\frac12 A_i\partial_jA_k-\frac{ig}{3}A_iA_jA_k\right).
 \label{eq:construct-cs}
\end{equation}
with $\epsilon^{123}=1$.
For the physical solution, we impose the usual Bunch--Davies boundary condition.

Let $A_\mu^{\mathrm{SD}}[a]$ be the full self-dual classical solution,\footnote{The self-dual solution serves only as a formal auxiliary reference; the Bunch--Davies condition is imposed on the physical solution.} satisfying
\begin{equation}
 F_{zi}[A^{\mathrm{SD}}[a]]
 +\frac12\epsilon_i{}^{jk}F_{jk}[A^{\mathrm{SD}}[a]]=0,
 \label{eq:construct-sd-equation}
\end{equation}
with the same boundary source as the full physical solution
$A_\mu^{\mathrm{phys}}[a]$,
\begin{equation}
 A_i^{\mathrm{SD}}[a](0,\bm x)=A_i^{\mathrm{phys}}[a](0,\bm x)
 =a_i(\bm x)=\sum_aJ_a\epsilon^+_{a,i}e^{i\bm k_a\cdot\bm x}.
 \label{eq:construct-common-source}
\end{equation}
The boundary source is linear in $J_a$, although the bulk solutions
 depend nonlinearly on these amplitudes. The Green functions used to
construct the two solutions differ by
\begin{equation}
 \displaystyle
 G_{\mathrm{SD}}^{ij}(z_1,z_2;\bm P)-G_{\mathrm{phys}}^{ij}(z_1,z_2;\bm P)
 =\frac{\epsilon_-^i(\bm P)\epsilon_+^j(\bm P)}{2p}
 (e^{-pz_1}-e^{pz_1})(e^{-pz_2}-e^{pz_2}),\qquad p=|\bm P|>0.
 \label{eq:construct-green-difference}
\end{equation}
Here $z_1,z_2$ are radial coordinates. 
The factorized difference in Eq.~\eqref{eq:construct-green-difference}
will relate the two tree expansions.

For $A^{\mathrm{SD}}[a]$, the self-duality equation~\eqref{eq:construct-sd-equation}
sets the bulk square in Eq.~\eqref{eq:construct-square} to zero, giving
\begin{equation}
 S[A^{\mathrm{SD}}[a]]=C_\partial[a].
\end{equation}
The tree-level generating functional is $\Psi_{\mathrm{SD}}[a]=-C_\partial[a]$.
Its cubic Chern--Simons term gives the all-plus three-point coefficient.
Since the boundary source $a_i$ is linear in $J_a$, there are no
higher-point contributions:
\begin{equation}
 \Psi_{\mathrm{SD},n}^{+\cdots+}
 =-\left.\frac{\partial^n C_\partial[a]}
 {\partial J_1\cdots\partial J_n}\right|_{J=0}
 =0,\qquad n\geq4.
 \label{eq:construct-chiral-zero}
\end{equation}

The factorized difference in Eq.~\eqref{eq:construct-green-difference}
supplies a negative-helicity leg at each endpoint, with each radial
factor implementing $D_p$ at fixed spatial momentum and polarization,
and an overall sewing weight $1/(2p)$, as illustrated in
Fig.~\ref{allplus:fig:n}(a).

Both fields solve the classical Yang--Mills equations, with their
perturbative solutions constructed using $G_{\mathrm{SD}}$ and
$G_{\mathrm{phys}}$, respectively. Expanding
$G_{\mathrm{SD}}=G_{\mathrm{phys}}+\Delta G$, terms without $\Delta G$
reproduce the physical coefficient $\Psi_n$. Each factor of $\Delta G$
cuts an internal line, and summing the remaining physical diagrams in
each component gives a full lower-point coefficient, including contact
interactions and longitudinal exchanges. At fixed color order, the
resulting sewing patterns correspond to noncrossing polygon dissections,
each counted once. Thus,
\begin{equation}
	\Psi_{\mathrm{SD},n}^{+\cdots+}
	=\Psi_n^{+\cdots+}
	+\sum_{\substack{C\ne\varnothing\\\mathrm{noncrossing}}}
	\left(\prod_{e\in C}\frac{1}{2p_e}\right)
	\prod_{f\in\mathrm{cells}(C)}
	\left[\left(\prod_{e\in C\cap\partial f}D_{e,f}\right)
	\Psi_{|f|}\right],\qquad n\geq4.
	\label{eq:construct-tree-expansion}
\end{equation}
Using $\Psi_{\mathrm{SD},n}^{+\cdots+}=0$ for $n\geq4$ and moving
the nonempty-cut terms to the other side gives the general
formula~\eqref{eq:ym-letter-alln}.

For gravity, the analogous decomposition holds for the renormalized
Einstein action evaluated on Einstein metrics~\cite{Anderson2001}.
We omit source-independent constants at fixed topology. The Hodge dual
uses the bulk orientation $dx^1\wedge dx^2\wedge dx^3\wedge dz$. With
$R_{\mu\nu}=-3L^{-2}g_{\mu\nu}$, $W^-=\tfrac12(W-{}^\star W)$,
and the common factor $M_{\rm Pl}^2L^2$ suppressed, one finds
\begin{equation}
 \begin{aligned}
 S_{\mathrm{EH,ren}}\big|_{\mathrm{Einstein}}
 &=-\frac18\int_M d^4x\,\sqrt g\,
 W_{\mu\nu\rho\sigma}W^{\mu\nu\rho\sigma}\\
 &=-\frac14\int_M d^4x\,\sqrt g\,
 W^-_{\mu\nu\rho\sigma}W^{-\,\mu\nu\rho\sigma}
 -\frac14\operatorname{CS}_{\Gamma}[g^{(0)}].
 \end{aligned}
 \label{eq:construct-gravity-square}
\end{equation}
Here
\begin{equation}
 \operatorname{CS}_{\Gamma}[g^{(0)}]
 =-\int d^3x\,\epsilon^{ijk}
 \left(\Gamma^a{}_{ib}\,\partial_j\Gamma^b{}_{ka}
 +\frac23\Gamma^a{}_{ib}\Gamma^b{}_{jc}\Gamma^c{}_{ka}\right),
\end{equation}
where $\Gamma$ is the Levi--Civita connection of $g^{(0)}=e^\gamma$.
On the self-dual Einstein reference solution, $W^-=0$, giving
\begin{equation}
 \Psi_{\mathrm{SD}}[\gamma]
 =-S_{\mathrm{EH,ren}}[g^{\mathrm{SD}}[\gamma]]
 =\frac14\operatorname{CS}_{\Gamma}[e^\gamma],
 \qquad
 \Psi_{\mathrm{SD},n}^{+\cdots+}
 =\frac14\operatorname{CS}_n^{+\cdots+}.
 \label{eq:construct-gravity-cs}
\end{equation}
Unlike gauge Chern--Simons, which is at most cubic in the boundary
source, the gravitational Chern--Simons functional depends nonlinearly
on $\gamma$ through the metric and its inverse in the connection.
Its source expansion therefore extends beyond cubic order, so
$\Psi_{\mathrm{SD},n}$ need not vanish for $n\geq4$. These local terms
account for the four- and five-point completions and motivate the
Chern--Simons contribution to the proposed all-multiplicity
relation~\eqref{eq:gr-letter-alln}.

\FloatBarrier
\section{Discussion}
\label{sec:discussion}

The all-plus wavefunction has a simpler analytic organization than
individual bulk diagrams suggest. In Yang--Mills theory, compatible
polygon dissections arrange the full lower-point coefficients into
sewn products with no additional uncut term. The absence of a generic
total-energy pole is then manifest, while the surviving partial-energy
poles and successive residues are determined by the factors in each
product. The associahedron describes both the compatible cuts and
the factorization of their combinatorial structure.

The proposed all-multiplicity gravity representation combines labelled
sewing trees with a local gravitational Chern--Simons completion. The
self-dual Einstein action explains the origin of this local term, which
supplies boundary data invisible to internal-energy cuts. Expanding the
boundary connection gives a systematic construction of the local kernels
at arbitrary multiplicity while keeping their three-derivative structure
explicit.

In gravity, the proposed all-multiplicity representation makes the
absence of total-energy poles manifest. This is stronger than the
vanishing of the all-plus flat-space amplitude, which fixes only the
coefficient of the leading total-energy singularity: all subleading
poles cancel as well. It would be very interesting to explore whether
an enhanced symmetry of the all-plus sector enforces these cancellations.


Both sewing constructions have a recursive structure in multiplicity, but they do not close within the all-plus sector: their building blocks are full lower-point wavefunctions with different helicity configurations. An arbitrary-multiplicity sewing relation therefore does not by itself provide an explicit all-\(n\) expression in terms of three-point data, unlike the all-multiplicity sQED result \cite{HeQED2026}. A natural next step is to extend the construction to the mixed-helicity sectors appearing in these building blocks and determine what additional input is required to obtain a closed recursive system.

\medskip
\noindent\begin{minipage}{\textwidth}
\acknowledgments
S.H. and Y.M. are supported by the National Natural Science Foundation
of China under Grant Nos.~12225510 and 12447101, and by the New
Cornerstone Science Foundation.
J.M. is supported by the European Union (ERC, UNIVERSE PLUS, 101118787).
Views and opinions expressed are, however, those of the authors only and
do not necessarily reflect those of the European Union or the European
Research Council Executive Agency. Neither the European Union nor the
granting authority can be held responsible for them.

All AI-assisted material and scientific conclusions were reviewed and
independently verified by the authors, who take full responsibility for
the content.
\end{minipage}

\FloatBarrier

\appendix
\section{Local gravitational contributions}
\label{app:local-gravity}

The Chern--Simons contribution is a local functional of the boundary
metric. We use the exponential source $g^{(0)}_{ij}=(e^\gamma)_{ij}$.
Its kernels $\operatorname{CS}_n$ are obtained by expanding the
gravitational Chern--Simons functional
\cite{MansiPetkouTagliabue2009,deHaro2009}
\begin{equation}
 \operatorname{CS}_{\Gamma}[g^{(0)}]
 =\int d^3x\,\epsilon^{ijk}
 \left(\Gamma^a{}_{ib}\,\partial_j\Gamma^b{}_{ka}
       +\frac23\Gamma^a{}_{ib}\Gamma^b{}_{jc}\Gamma^c{}_{ka}\right),
 \label{eq:gr-cs-functional}
\end{equation}
where $\Gamma$ is the Levi--Civita connection of $g^{(0)}$.
The orientation and overall normalization of the kernels are fixed
by the four- and five-point expressions below; the contribution to
$\Psi_n$ is $\tfrac14\operatorname{CS}_n$ in our convention.

The nonlinear dependence of $\Gamma$ on $\gamma$ produces
a local boundary kernel at every multiplicity. These kernels contain
only boundary spatial derivatives and introduce no total- or
partial-energy poles.

\subsection{Connection expansion and coefficient extraction}
\label{app:cs-generator}

{We first keep the boundary polarizations general and write}
\begin{equation}
 {\gamma_{ij}(\bm x)=\sum_{a=1}^n
 J_a\,h_{a,ij}\,e^{i\bm k_a\cdot\bm x},}
 \label{eq:cs-general-source}
\end{equation}
{where each $h_a$ is a symmetric matrix.}

{
We use the matrix exponential
\[
 e^\gamma=\sum_{m=0}^\infty\frac{\gamma^m}{m!}
 =\mathbf1+\gamma+\frac{\gamma^2}{2!}+\frac{\gamma^3}{3!}+\cdots,
 \qquad g^{(0)}=e^\gamma.
\]
The powers are matrix products, for example
$(\gamma^2)_{ij}=\gamma_i{}^k\gamma_{kj}$;
the inverse metric is $e^{-\gamma}$.
}

The expansion can be performed before choosing helicities. Regard
$\gamma_i{}^j$ as a matrix with indices raised by the flat boundary
metric, introduce a bookkeeping parameter $t$, and write the
Levi--Civita connection one-form as
$\Gamma[e^{t\gamma}]=\sum_{r\geq1}t^r\Gamma_r$.
Since $g^{-1}=e^{-t\gamma}$, its homogeneous components are
\begin{equation}
	(\Gamma_r)^a{}_b
	=\frac12\sum_{u=0}^{r-1}\frac{(-1)^u}{u!(r-u)!}
	(\gamma^u)^{ac}
	\left[\partial_i(\gamma^{r-u})_{cb}
	+\partial_b(\gamma^{r-u})_{ci}
	-\partial_c(\gamma^{r-u})_{ib}\right]dx^i .
	\label{eq:cs-connection-generator}
\end{equation}
The derivative acts on the entire matrix power, with the order of
$\gamma$ and its derivatives kept as displayed. Define the degree-$n$ functional by
$\operatorname{CS}_\Gamma[e^{t\gamma}]=\sum_{n\geq2}t^n\mathcal I_n[\gamma]$.
Substitution in Eq.~\eqref{eq:gr-cs-functional} gives
\begin{equation}
	\mathcal I_n=\int\operatorname{tr}\left[
	\sum_{r=1}^{n-1}\Gamma_r\wedge d\Gamma_{n-r}
	+\frac23\!\sum_{\substack{r,s,\ell\geq1\\r+s+\ell=n}}
	\Gamma_r\wedge\Gamma_s\wedge\Gamma_\ell\right].
	\label{eq:cs-alln-generator}
\end{equation}

{
The single parameter $t$ counts the total field degree, so after
Eq.~\eqref{eq:cs-general-source}, $\mathcal I_n$ has total degree $n$
in the $J_a$. The $J_a$ distinguish individual legs: the labelled kernel
extracts the mixed monomial $J_1\cdots J_n$, with the overall momentum
delta function removed,
\[
 \operatorname{CS}_n(h_1,\ldots,h_n)
 =\left.\frac{\partial^n\mathcal I_n[\gamma]}
 {\partial J_1\cdots\partial J_n}\right|_{J=0}.
\]
There is no additional $n!$.
}
Spatial derivatives become external
momenta, and every term retains the three derivatives of the original
functional.

\subsection{Four- to seven-point expansions}

The four- to seven-point expressions have a short form at the level of
connection tensors. Below, products of one-forms denote wedge products
with matrix multiplication in the displayed order. Cyclicity of the
trace and spatial integration by parts give

\begingroup
\begin{align}
 \mathcal I_4=\int\operatorname{tr}\bigl[&
 2\Gamma_1d\Gamma_3+\Gamma_2d\Gamma_2
 +2\Gamma_1\Gamma_1\Gamma_2\bigr],
 \label{eq:cs-four-compact}\\
 \mathcal I_5=2\int\operatorname{tr}\bigl[&
 \Gamma_1d\Gamma_4+\Gamma_2d\Gamma_3
 +\Gamma_1\Gamma_1\Gamma_3+\Gamma_1\Gamma_2\Gamma_2\bigr].
 \label{eq:cs-five-compact}
\end{align}
\endgroup

\begin{align}
 \mathcal I_6=\int\operatorname{tr}\bigl[&
 2\Gamma_1d\Gamma_5+2\Gamma_2d\Gamma_4+\Gamma_3d\Gamma_3
 +2\Gamma_1\Gamma_1\Gamma_4\nonumber\\*
 &+2\Gamma_1\Gamma_2\Gamma_3+2\Gamma_1\Gamma_3\Gamma_2
 +\tfrac23\Gamma_2\Gamma_2\Gamma_2\bigr],
 \label{eq:cs-six-compact}\\
 \mathcal I_7=2\int\operatorname{tr}\bigl[&
 \Gamma_1d\Gamma_6+\Gamma_2d\Gamma_5+\Gamma_3d\Gamma_4
 +\Gamma_1\Gamma_1\Gamma_5\nonumber\\*
 &+\Gamma_1\Gamma_2\Gamma_4+\Gamma_1\Gamma_4\Gamma_2
 +\Gamma_1\Gamma_3\Gamma_3+\Gamma_2\Gamma_2\Gamma_3\bigr].
 \label{eq:cs-seven-compact}
\end{align}

Here equality is after spatial integration, with no boundary
contribution at spatial infinity (or, equivalently, after imposing
momentum conservation on the labelled kernel). For example,
$\int\operatorname{tr}(\Gamma_r d\Gamma_s)
=\int\operatorname{tr}(\Gamma_s d\Gamma_r)$, while the cyclic
permutations of three one-forms have positive graded sign. These two
facts account for every coefficient in
Eqs.~\eqref{eq:cs-four-compact}--\eqref{eq:cs-seven-compact}.

\subsection{All-plus helicity kernels}

{
For the all-plus kernels $\operatorname{CS}_n^{+\cdots+}$, set
$h_{a,ij}=\xi^+_{a,ij}=\epsilon^+_{a,i}\epsilon^+_{a,j}$ and write
the source as}
\begin{equation}
 \gamma_{ij}(\bm x)=\sum_{a=1}^n
 J_a\,\xi^+_{a,ij}\,e^{i\bm k_a\cdot\bm x},
 \label{eq:gr-cs-labelled-source}
\end{equation}
{
Insert this Fourier expansion into $\mathcal I_n$, take the coefficient
of $J_1\cdots J_n$, and perform the spatial integral. Removing the
overall factor $(2\pi)^3\delta^3(\sum_a\bm k_a)$ leaves the kernel
$\operatorname{CS}_n^{+\cdots+}$.
}

{
To express these kernels in dS spinor-helicity form, we use
$k_i=-\tfrac12\langle i\bar i\rangle$ and the following
relations~\cite{MaldacenaPimentel2011,BaumannSpinning2021}:}
\begin{equation}
\begin{gathered}
 \langle i j\rangle
   =\epsilon_{\alpha\beta}\lambda_i^\alpha\lambda_j^\beta,
 \qquad
 \langle i\bar j\rangle
   =\epsilon_{\alpha\beta}\lambda_i^\alpha\bar\lambda_j^\beta,
 \qquad
 \langle\bar i\bar j\rangle
   =\epsilon_{\alpha\beta}\bar\lambda_i^\alpha\bar\lambda_j^\beta,
 \\[4pt]
 {
 \epsilon_i^+\cdot\epsilon_j^+=-\frac{\langle\bar i\bar j\rangle^2}{8k_i k_j},
 \qquad
 \epsilon_i^+\cdot\bm k_j
 =\frac{\langle j\bar i\rangle\langle\bar i\bar j\rangle}{4k_i}.}
\end{gathered}
 \label{eq:cs-dot-spinor-conversion}
\end{equation}

{
Applying this procedure to $\mathcal I_4$ and $\mathcal I_5$ in
Eqs.~\eqref{eq:cs-four-compact}--\eqref{eq:cs-five-compact} gives
the following four- and five-point kernels, with no total- or
partial-energy poles:}
\begin{equation}
 \operatorname{CS}_4^{++++}
 =\frac{\sum_{\sigma\in S_4}F_4(\sigma)}
 {4096\prod_{a=1}^4 k_a^2},\qquad
 \operatorname{CS}_5^{+++++}
 =\frac{\sum_{\sigma\in S_5}F_5(\sigma)}
 {65536\prod_{a=1}^5 k_a^2},
 \label{eq:gr-short-cs-sum}
\end{equation}
where the permutation sums are not averaged. A compact choice of seeds is
\begin{equation}
\begin{aligned}
 &F_4(1,2,3,4)=\frac{k_1\langle\bar1\bar2\rangle^2}{24}\Big\{
 2k_1(9k_2-4k_1)
 \langle\bar1\bar3\rangle^2\langle\bar2\bar4\rangle^2\langle\bar3\bar4\rangle^2
 +3\langle\bar1\bar3\rangle^2\langle\bar2\bar4\rangle^2
 \Big(3\langle1\bar3\rangle\langle\bar3\bar1\rangle
          \langle3\bar4\rangle\langle\bar4\bar3\rangle
 \\
 &\qquad-2\langle2\bar3\rangle\langle\bar3\bar2\rangle
          \langle2\bar4\rangle\langle\bar4\bar2\rangle\Big)
 +3\langle\bar2\bar3\rangle^2\langle2\bar4\rangle\langle\bar4\bar2\rangle
 \Big(5\langle\bar1\bar3\rangle^2\langle2\bar4\rangle\langle\bar4\bar2\rangle
      -2\langle\bar3\bar4\rangle^2\langle2\bar1\rangle\langle\bar1\bar2\rangle\Big)
 \Big\},
\end{aligned}
 \label{eq:gr-short-cs4}
\end{equation}
\begin{equation}
\begin{aligned}
 &F_5(1,2,3,4,5)=
 \frac{\langle\bar1\bar2\rangle^4\langle\bar3\bar4\rangle^2
            \langle\bar3\bar5\rangle^2}{24}
 \Big[3k_4\langle3\bar4\rangle\langle\bar4\bar3\rangle
              \langle4\bar5\rangle\langle\bar5\bar4\rangle
 -k_3\langle3\bar4\rangle\langle\bar4\bar3\rangle
              \langle3\bar5\rangle\langle\bar5\bar3\rangle
 \\
 &\quad+2k_2\langle1\bar4\rangle\langle\bar4\bar1\rangle
              \langle4\bar5\rangle\langle\bar5\bar4\rangle
-6k_1\langle\bar4\bar5\rangle^2
       \Big(4k_3k_4-\tfrac12\langle34\rangle\langle\bar3\bar4\rangle\Big)\Big]
 \\[4pt]
 &\quad+\frac{\langle\bar1\bar2\rangle^2\langle\bar1\bar3\rangle^2
              \langle\bar2\bar3\rangle^2\langle\bar4\bar5\rangle^2}{12}
       \langle1\bar4\rangle\langle\bar4\bar1\rangle
 \Big(5k_2\langle1\bar5\rangle\langle\bar5\bar1\rangle
                     -k_1\langle2\bar5\rangle\langle\bar5\bar2\rangle\Big)
 \\[4pt]
 &\quad+\frac{\langle\bar1\bar2\rangle^2\langle\bar1\bar3\rangle^2
              \langle\bar2\bar4\rangle^2\langle\bar3\bar5\rangle^2}{12}
 \Big[2k_3\langle2\bar4\rangle\langle\bar4\bar2\rangle
              \langle4\bar5\rangle\langle\bar5\bar4\rangle
 +5k_5\langle1\bar4\rangle\langle\bar4\bar1\rangle
              \langle2\bar5\rangle\langle\bar5\bar2\rangle
       -6k_1k_4^2\langle\bar4\bar5\rangle^2
 \\
 &\qquad+\frac{\langle1\bar4\rangle\langle\bar4\bar1\rangle}{k_1}
       \Big\{\Big(k_2k_3-\tfrac12\langle23\rangle\langle\bar2\bar3\rangle\Big)
                    \langle4\bar5\rangle\langle\bar5\bar4\rangle-\Big(k_3k_4-\tfrac12\langle34\rangle\langle\bar3\bar4\rangle\Big)
                    \langle2\bar5\rangle\langle\bar5\bar2\rangle\Big\}\Big].
\end{aligned}
 \label{eq:gr-short-cs5}
\end{equation}


\providecommand{\href}[2]{#2}
\begingroup\raggedright

\endgroup

\end{DocumentChineseNames}
\end{document}